\documentclass[amsfonts,aps,nofootinbib,preprint,superscriptaddress]{revtex4}

\usepackage{graphicx}

\begin{document}

\title{The Linear Delta Expansion applied to the O'Raifeartaigh model}

\author{M. C. B. Abdalla{\footnote{mabdalla@ift.unesp.br}}}
\affiliation{Instituto de F\'{\i}sica Te\'{o}rica, UNESP, Rua Pamplona
145, Bela Vista, S\~{a}o Paulo, SP, 01405-900, Brazil }
\author{J. A. Helay\"{e}l-Neto{\footnote{helayel@cbpf.br}}}
\affiliation{Centro Brasileiro de Pesquisas F\'isicas, Rua Dr. Xavier Sigaud 150, 
Urca, Rio de Janeiro, RJ, 22290-180, Brazil}
\author{Daniel L. Nedel{\footnote{dnedel.unipampa@ufpel.tche.br}}}
\affiliation{Universidade Federal do Pampa, 
Rua Carlos Barbosa S/N, Bairro Get\'ulio Vargas, 96412-420, Bag\'e, RS, Brazil}
\author{Carlos R. Senise Jr.{\footnote{senise@ift.unesp.br}}}
\affiliation{Instituto de F\'{\i}sica Te\'{o}rica, UNESP, Rua Pamplona
145, Bela Vista, S\~{a}o Paulo, SP, 01405-900, Brazil }

\begin{abstract}
We reassess the method of the linear delta expansion for the calculation of effective potentials in superspace, by adopting the improved version of the super-Feynman rules in the framework of the O'Raifeartaigh model for spontaneous supersymmetry breaking. The effective potential is calculated using both the fastest apparent convergence and the principle of minimal sensitivity criteria and the consistency and efficacy of the method are checked in deriving the Coleman-Weinberg potential.
\end{abstract}

\maketitle

O'Raifeartaigh-type models for spontaneous breaking of supersymmetry (SUSY) have recently received renewed attention. %This is because 
According to the Nelson-Seiberg theorem \cite{NS}, all these models have an \textit{R} symmetry, which plays an important role in SUSY breaking. However, in order to have nonzero Majorana gaugino masses, \textit{R} symmetry needs to be broken \cite{IS}. Since the simplest original O'Raifeartaigh model \cite{O'Raifeartaigh} does not spontaneously break \textit{R} symmetry, generalized O'Raifeartaigh models, which spontaneously violate \textit{R} symmetry, have been constructed \cite{IS,ISS1,ISS2}.

In many generalized O'Raifeartaigh models, \textit{R} symmetry is spontaneously broken by the pseudomoduli, which are charged under \textit{R} symmetry and acquire a nonzero vacuum expectation value via effective potential. Using this approach, it was shown how to build up models that break \textit{R} symmetry at one-loop via the Coleman-Weinberg potential \cite{Shih,Marques}.
 
Since the Coleman-Weinberg potential \cite{CW} is a sum of all one-loop diagrams of the theory, it is very interesting to develop methods that account for higher loop corrections in the effective potential of O'Raifeartaigh-type models and go beyond the approximation used in \cite{Shih}. There are two traditional ways to make resummations in supersymmetric and nonsupersymmetric quantum field theories: the diagrammatic calculation and the functional calculation \cite{CW,Jac}. Both are very difficult to use if we are interested in going beyond the Coleman-Weinberg approximation. Mainly because it is necessary to work with infinite diagrams, which turns the renormalization procedure into a heavy task. 
     
Over the past years, an alternative resummation method has been developed, namely, the linear delta expansion (LDE) \cite{delta1}. This method can easily reproduce the Coleman-Weinberg potential, and the use of the LDE in various quantum field theory models has proven to be a powerful tool to derive new nonperturbative results \cite{Kneur,Ricardo}. In \cite{nos}, the method was further developed to be applied to supersymmetric theories in superspace, where the Coleman-Weinberg potential has been derived and two-loop corrections for the K\"{a}hler potential of the Wess-Zumino model have been computed. The main characteristic of the method is to use a traditional perturbative approach together with an optimization procedure. So, in order to derive a nonperturbative result, it is just necessary to work with a few diagrams and use perturbative renormalization techniques.

The main goal of this paper is to show that the LDE can also be a powerful method to derive nonperturbative results in O'Raifeartaigh-like models, which go beyond the Coleman-Weinberg approximation. This is very important in order to understand the effects of nonperturbative corrections to SUSY and \textit{R} symmetry breaking. To this end, in Sec. I, we present the main steps of the method based on the LDE in superspace; in Sec. II, we further develop the method to be applied in the O'Raifeartaigh model. We also show that the method induces soft breaking terms in the Lagrangian. In Sec. III, we calculate the effective potential by adopting the fastest apparent convergence criterion; in Sec. IV we show that the same solution is derived using the principle of minimal sensitivity criterion. Concluding remarks are finally cast in Sec. V. In the Appendix, we present the detailed calculation of the vacuum diagram in superspace and derive the Coleman-Weinberg potential presented in Sec. III.

\section{The Linear Delta Expansion in Superspace}
In this section, we make a brief review of the LDE. Starting with a Lagrangian ${\cal L}$, let us define the following interpolated Lagrangian ${\cal L}^{\delta}$:
\begin{equation}
{\mathcal L}^{\delta}=\delta{\mathcal L}(\mu)+(1-\delta){\mathcal L}_{0}(\mu) \ , \label{LDE} 
\end{equation}
where $\delta$ is an arbitrary parameter, ${\cal L}_0(\mu)$ is the free Lagrangian, and $\mu$ is a mass parameter. Note that, when $\delta =1$, the original theory is retrieved. The $\delta$ parameter labels interactions and it is used as a perturbative coupling instead of the original coupling. The mass parameter appears in ${\cal L}_0$ and $\delta{\cal L}_0$. In fact, we are using the traditional trick that consists in adding and subtracting a mass term in the original Lagrangian. The $\mu$-dependence of ${\cal L}_0$ is absorbed into the propagators, whereas $\delta{\cal L}_0$ is regarded as a quadratic interaction. 
 
Let us now define the strategy of the method. We apply a usual perturbative expansion in $\delta$, and at the end, we set $\delta =1$. Up to this stage, traditional perturbation theory is applied, working with finite Feynman diagrams, and the results are purely perturbative. However, quantities evaluated at finite order in $\delta$ explicitly depend on $\mu$. So it is necessary to fix the $\mu$ parameter. There are two ways to do that. The first one is to use the principle of minimal sensitivity (PMS) \cite{PMS}. Since $\mu$ does not belong to the original theory, we may require that a physical quantity, such as the effective potential $V^{(k)}(\mu)$, calculated perturbatively to order $\delta ^k$, must be evaluated at a point where it is less sensitive to the parameter $\mu$. According to the PMS, $\mu={\mu_0}$ is the solution to the equation
\begin{equation}
\left.\frac{\partial V^{(k)}(\mu)}{\partial\mu}\right|_{\mu=\mu_0,\delta=1}=0 \ . \label{PMS}
\end{equation}
After this procedure, the optimum value, ${\mu_0}$, will be a function of the original coupling and fields. Then, we replace ${\mu_0}$ into the effective potential $V^{(k)}$ and obtain a nonperturbative result, since the propagator depends on $\mu$. 
 
The second way to fix $\mu$ is known as the fastest apparent convergence (FAC) criterion \cite{PMS}. It requires that, from any $k$ coefficient of the perturbative expansion
 
\begin{equation}
V^{(k)}(\mu)=\sum_{i=0}^{k}c_i(\mu)\delta^i \label{coef} \ ,
\end{equation}
the following relation be fulfilled:   
 
\begin{equation}
\left.\left[V^{(k)}(\mu)- V^{(k-1)}(\mu)\right]\right|_{\delta=1}=0 \ . \label{fac}
\end{equation}
 
Again, the ${\mu_0}$ solution of the above equation will be a function of the original couplings and fields, and whenever we replace $\mu={\mu_0}$ into $V(\mu)$, we obtain a nonperturbative result. Equation (\ref{fac}) is equivalent to taking the $k$th coefficient of (\ref{coef}) equal to zero ($c_k=0$). If we are interested in an order-$\delta^{k}$ result [$V^{(k)}(\mu)$] using the FAC criterion, it is just necessary to find the solution to the equation $\displaystyle\left.c_{k+1}(\mu)\right|_{\mu=\mu_0}=0$ and plug it into $V^{(k)}(\mu)$. Reference \cite{Kneur} provides an extensive list of successful applications of the method. 

Let us now further develop the LDE for superspace applications. Following Ref. \cite{nos}, for general models with chiral and antichiral superfields, we need to implement two mass parameters, $\mu$ and $\bar{\mu}$, instead of just one. In order to fix these parameters, we employ two optimization equations. In particular, we use the FAC criterion, so that we have one superspace equation similar to (\ref{fac}). Also, we need to take care of the vacuum diagrams. In general, when the effective potential is calculated in quantum field theory, we do not worry about vacuum diagrams, since they do not depend on fields. However, the vacuum diagrams depend on $\mu$ and are important to the LDE, since the arbitrary mass parameter will depend on fields after the optimization procedure. So, in the LDE, it is necessary to calculate the vacuum diagrams order by order. On the other hand, it is well-known that, in superspace, vacuum superdiagrams are identically zero, by virtue of Berezin integrals. To avoid this, we have to consider, from the very beginning, the parameters $\mu$, $\bar{\mu}$ as superfields and keep the vacuum supergraphs until the optimization procedure is carried out. In order to make the procedure clear, let us write the interpolated Lagrangian, ${\cal L}^\delta$, for the Wess-Zumino model discussed in \cite{nos}:
\begin{eqnarray}
{\mathcal L}^{\delta}&=&\delta{\mathcal L}(\mu,\bar{\mu})+(1-\delta){\mathcal L}_{0}(\mu,\bar{\mu}) \nonumber\\
&=&\int\!d^{4}\theta\bar{\phi}\phi+\!\int\!d^{2}\theta\left(\frac{M}{2}\phi^{2}+\frac{\delta\lambda}{3!}\phi^{3}-\frac{\delta\mu}{2}\phi^{2}\right)+\!\int\!d^{2}\bar{\theta}\left(\frac{\bar{M}}{2}\bar{\phi}^{2}+\frac{\delta\bar{\lambda}}{3!}\bar{\phi}^{3}-\frac{\delta\bar{\mu}}{2}\bar{\phi}^{2}\right),  \label{linter}
\end{eqnarray} 
where $m$ is the original mass, $M=m+\mu$ and $\bar{M}=m+\bar{\mu}$. Now, one has a new chiral and antichiral quadratic interaction proportional to $\delta\mu$ and $\delta\bar{\mu}$. Also the superpropagator will have a dependence on $\mu$ and $\bar{\mu}$. In the O'Raifeartaigh model, we shall expand $\mu$ and $\bar{\mu}$ as chiral and antichiral superfields. This procedure will generate spurion interactions in the Lagrangian, which will explicitly break supersymmetry.   

From the generating superfunctional in the presence of the chiral ($J$) and antichiral ($\bar J$) sources
\begin{equation}
\tilde{Z}[J,\bar{J}]=exp\left[iS_{INT}\left(\frac{1}{i}\frac{\delta}{\delta J},\frac{1}{i}\frac{\delta}{\delta\bar{J}}\right)\right]exp\left[\frac{i}{2}
(J,\bar{J})G^{(M,\bar{M})}\left( \begin{array}{c}
J \\
\bar{J}
\end{array}
\right)\right] \ ,
\end{equation}
we can write the supereffective action:
\begin{equation}
\Gamma[\phi,\bar{\phi}]=-\frac{i}{2}\ln[sDet\left(G^{(M,\bar{M})}\right)]-i\ln\tilde{Z}[J,\bar{J}]-\int\!d^{6}zJ(z)\phi(z)-\int\!d^{6}\bar{z}\bar{J}(z)\bar{\phi}(z),             \label{sea}
\end{equation}
where $G^{(M,\bar{M})}$ is the matrix propagator and $sDet\left(G^{(M,\bar{M})}\right)$ is the superdeterminant of $G^{(M,\bar{M})}$, which, in general, is equal to one; but here we keep it, because $G^{(M,\bar{M})}$ depends on $\mu$ and $\bar{\mu}$. Also, due to the $\mu$ and $\bar{\mu}$ dependence, the supergenerator of the vacuum diagrams, $\tilde{Z}[0,0]$, is not identically equal to one. We can define the normalized functional generator as $Z_N = \frac{\tilde{Z}[J,\bar J]}{\tilde{Z}[0,0]}$, and write the effective action as
\begin{equation}
\Gamma[\phi,\bar{\phi}]=-\frac{i}{2}\ln[sDet(G)]-i\ln\tilde{Z}[J_{0},\bar{J}_{0}]+\Gamma_{N}[\phi,\bar{\phi}] \ , \label{Gamma Phi barPhi}
\end{equation}
where the sources $J_0$ and $\bar{J}_0$ are defined by the equations 
\begin{eqnarray}
\frac{\delta W[J,\bar{J}]}{\delta J(z)}|_{J=J_{0}}=\frac{\delta W[J,\bar{J}]}{\delta \bar{J}(z)}|_{\bar{J}=\bar{J}_{0}}=\frac{\delta\tilde{Z}[J,\bar{J}]}{\delta J(z)}|_{J=J_{0}}=\frac{\delta\tilde{Z}[J,\bar{J}]}{\delta\bar{J}(z)}|_{\bar{J}=\bar{J}_{0}}=0 \ . \label{gerador}
\end{eqnarray}
In (\ref{Gamma Phi barPhi}), the first two terms represent the vacuum diagrams (which are usually zero) and $\Gamma_N[\phi,\bar{\phi}]$ is the usual contribution to the effective action.

\section{LDE in the O'Raifeartaigh model}
Let us now derive the interpolated Lagrangian and the new Feynman rules for the O'Raifeartaigh model. 
 
The simplest O'Raifeartaigh model is described by the following Lagrangian:
 
\begin{equation}
{\mathcal L}=\int d^{4}\theta\bar{\phi}_{i}\phi_{i}-\left[\int d^{2}\theta\left(\xi\phi_{0}+m\phi_{1}\phi_{2}+g\phi_{0}\phi_{1}^{2}\right)+h.c.\right] \ , \label{O'Raifeartaigh}
\end{equation} 
where $i=0,1,2$.
  
In order to take into account the nonperturbative contributions of all fields in the model, we need to implement the LDE with the matrix mass parameters $\mu_{ij}$ and $\bar{\mu}_{ij}$. Adding and subtracting these mass terms in the Lagrangian of a general O'Raifeartaigh model we obtain
\begin{equation}
{\mathcal L}(\mu,\bar{\mu})={\mathcal L}_{0}(\mu,\bar{\mu})+{\mathcal L}_{int}(\mu,\bar{\mu}) \ ,
\end{equation}
where
\begin{eqnarray} 
{\mathcal L}_{0}(\mu,\bar{\mu})&=&\int d^{4}\theta\bar{\phi}_{i}\phi_{i}-\left[\int d^{2}\theta\left(\xi_{i}\phi_{i}+\frac{1}{2}M_{ij}\phi_{i}\phi_{j}\right)+h.c.\right] \ , \\
{\mathcal L}_{int}(\mu,\bar{\mu})&=&-\left[\int d^{2}\theta\left(\frac{1}{3!}g_{ijk}\phi_{i}\phi_{j}\phi_{k}-\frac{1}{2}\mu_{ij}\phi_{i}\phi_{j}\right)+h.c.\right] \ ,  
\end{eqnarray}
with $M_{ij}=m_{ij}+ \mu_{ij}$ and $i,j,k=0,1,2$ are symmetrical indices.

Generally, when SUSY breaking is studied in superspace, soft explicit breaking terms naturally arise. The latter have been carefully classified and studied by Girardello and Grisaru \cite{Girardello-Grisaru}. Here, soft breaking terms automatically appear from the $\mu$ and $\bar{\mu}$ dependence. Let us expand the arbitrary mass parameters as chiral and antichiral superfields:
\begin{equation}
\mu_{ij}=\lambda_{ijk}\varphi_k=\lambda_{ijk}(\rho_k+\theta^2\chi_k)=\lambda_{ijk}\rho_k+\lambda_{ijk}\chi_k\theta^2=\rho_{ij}+b_{ij}\theta^2 \ ,
\end{equation}
so that
\begin{equation}
M_{ij}=m_{ij}+\mu_{ij}=(m_{ij}+\rho_{ij})+b_{ij}\theta^2=a_{ij}+b_{ij}\theta^2 \ .
\end{equation}

Now, the interpolated Lagrangian (\ref{LDE}) becomes
\begin{equation}
{\mathcal L}^{\delta}= {\mathcal L}_{0}^{\delta} + {\mathcal L}_{int}^{\delta} \ ,
\end{equation}
where the free Lagrangian, ${\mathcal L}^{\delta}_0$, is 
\begin{equation}
{\mathcal L}_{0}^{\delta}=\int d^{4}\theta\bar{\phi}_{i}\phi_{i}-\left[\int d^{2}\theta\left(\xi_{i}\phi_{i}+\frac{1}{2}a_{ij}\phi_{i}\phi_{j}+\frac{1}{2}b_{ij}\theta^{2}\phi_{i}\phi_{j}\right)+h.c.\right] \ ,
\end{equation}
and the interaction Lagrangian reads as follows:
\begin{equation}
{\mathcal L}_{int}^{\delta}=-\left[\int d^{2}\theta\left(\frac{\delta}{3!}g_{ijk}\phi_{i}\phi_{j}\phi_{k}-\frac{\delta}{2}\mu_{ij}\phi_{i}\phi_{j}\right)+h.c.\right] \ .
\end{equation}

Notice that the interaction Lagrangian has now soft breaking terms proportional to the $\mu$ components. We are going to treat these terms perturbatively in $\delta$, like all interactions. Clearly, the method does not change the renormalization aspects of the theory \footnote{See Ref. \cite{Nibbelink} for a discussion on the renormalization of softly broken SUSY.}. 

Now, in order to get the simplest O'Raifeartaigh model when $\delta=1$ (\ref{O'Raifeartaigh}), we make the choices
\begin{equation}
\left\{\begin{array}{lllll}
          \xi_{0}=\xi \ ; \\
          M_{01}=a_{01}=\rho_{01}=a \ ; \\
          M_{11}=b_{11}\theta^{2}=b\theta^{2} \ ; \\
          M_{12}=a_{12}=m_{12}+\rho_{12}=m+\rho=M \ ; \\
          g_{011}=g \ ,
          \end{array}\right. \label{choice}
\end{equation}
and all other $\xi_{i}$ and $M_{ij}$ set to zero. With that, we obtain
\begin{eqnarray}
{\mathcal L}_{0}^{\delta}&=&\int d^{4}\theta\bar{\phi}_{i}\phi_{i}-\left[\int d^{2}\theta\left(\xi\phi_{0}+M\phi_{1}\phi_{2}+a\phi_{0}\phi_{1}+\frac{1}{2}b\theta^{2}\phi_{1}^{2}\right)+h.c.\right] \ , \nonumber\\
{\mathcal L}_{int}^{\delta}&=&-\left[\int d^{2}\theta\left(\delta g\phi_{0}\phi_{1}^{2}-\delta\rho\phi_{1}\phi_{2}-\delta a\phi_{0}\phi_{1}-\frac{\delta}{2}b\theta^{2}\phi_{1}^{2}\right)+h.c.\right] \ .
\end{eqnarray}

The new propagators can be derived from the free Lagrangian, which also has explicit dependence on $\theta$ and $\bar{\theta}$ from the $\mu$ and $\bar{\mu}$ components. The basic techniques for dealing with such an explicit dependence were developed a long time ago in \cite{Helayel}. Using these techniques, the new propagators can be written as (here we write only the $\phi_{1}\phi_{1}$, $\phi_{0}\phi_{1}$ and $\phi_{1}\phi_{2}$ propagators, since only these appear in the order-$\delta^1$ effective potential) 
\begin{eqnarray}
\langle T(\phi_{1}\phi_{1})\rangle&=&\frac{\bar{b}}{\left[(k^{2}+\left|M\right|^{2}+|a|^{2})^{2}-|b|^{2}\right]}\frac{1}{4}\bar{\theta}_{1}^{2}D_{1}^{2}\delta^{4}(\theta_{1}-\theta_{2}) \ , \\
\langle T(\phi_{0}\phi_{1})\rangle&=&\frac{\bar{a}}{\left[k^{2}(k^{2}+\left|M\right|^{2}+|a|^{2})\right]}\frac{1}{4}D_{1}^{2}\delta^{4}(\theta_{1}-\theta_{2}) \nonumber\\
& \  &-\frac{\bar{a}|b|^{2}}{(k^{2}+\left|M\right|^{2}+|a|^{2})\left[(k^{2}+\left|M\right|^{2}+|a|^{2})^{2}-|b|^{2}\right]}\frac{1}{4}\theta_{1}^{2}\bar{\theta}_{1}^{2}D_{1}^{2}\delta^{4}(\theta_{1}-\theta_{2}) \ , \\
\langle T(\phi_{1}\phi_{2})\rangle&=&\frac{\bar{M}}{\left[k^{2}(k^{2}+\left|M\right|^{2}+|a|^{2})\right]}\frac{1}{4}D_{1}^{2}\delta^{4}(\theta_{1}-\theta_{2}) \nonumber\\
& \  &-\frac{\bar{M}|b|^{2}}{(k^{2}+\left|M\right|^{2}+|a|^{2})\left[(k^{2}+\left|M\right|^{2}+|a|^{2})^{2}-|b|^{2}\right]}\frac{1}{4}D_{1}^{2}\theta_{1}^{2}\bar{\theta}_{1}^{2}\delta^{4}(\theta_{1}-\theta_{2}) \ .
\end{eqnarray}
Also, we can write the new Feynman rules for the vertices:
\begin{eqnarray}
\phi_{0}\phi_{1}^{2} \ \hbox{vertex}&:&-2\delta g\int d^{4}\theta \ ; \nonumber\\
\phi_{1}\phi_{2} \ \hbox{vertex}&:&\delta\rho\int d^{4}\theta \ ; \\
\phi_{0}\phi_{1} \ \hbox{vertex}&:&\delta a\int d^{4}\theta \ ; \nonumber\\
\phi_{1}\phi_{1} \ \hbox{vertex}&:&\delta b\int d^{4}\theta\theta^{2} \ . \nonumber \label{vértices}
\end{eqnarray}

In order to calculate the first term of the effective action defined by (\ref{Gamma Phi barPhi}), one can adopt the same techniques as in \cite{Nibbelink}, and write this term as a supertrace:
\begin{equation} 
{\mathcal V}_{eff}^{(0)}=-\frac{1}{2}\int d^{4}\theta_{12}d^{4}\delta_{21}Tr\ln[P^{T}K]\delta_{21} \ , \label{ordem0}
\end{equation}
where $d^{4}\theta_{12}=d^{4}\theta_{1}d^{4}\theta_{2}$, $\delta_{21}=\delta^{4}(\theta_2-\theta_1)$, and the notation $Tr$ refers to the trace over the chiral multiplets in the real basis defined by the vector $(\Phi^T,\bar{\Phi})^T$. $P$ is the matrix defined by the chiral projectors $P_{+}=\frac{\bar{D}^{2}D^{2}}{16\Box}$ and $P_{-}=\frac{D^{2}\bar{D}^{2}}{16\Box}$ as
\begin{equation}
P=\left(\begin{array}{cc}
          0 & P_{-} \\
          P_{+} & 0
          \end{array}\right) \ , \label{P}
\end{equation}
and 
\begin{equation}
K=\left(\begin{array}{cc}
          \left(AP_{-}+B\frac{1}{\Box^{1/2}}\eta_{-}\right)\frac{D^2}{4\Box} & \textbf{1}_{3\times3} \\
          \textbf{1}_{3\times3} & \left(\bar{A}P_{+}+\bar{B}\frac{1}{\Box^{1/2}}\bar{\eta}_{+}\right)\frac{\bar{D}^2}{4\Box}
          \end{array}\right) \ , \label{K}
\end{equation}
with
\begin{equation}
A=\left(\begin{array}{ccc}
          0 & a & 0 \\
          a & 0 & M \\
          0 & M & 0
          \end{array}\right) \ \ , \ \ 
B=\left(\begin{array}{ccc}
          0 & 0 & 0 \\
          0 & b & 0 \\
          0 & 0 & 0
          \end{array}\right) \ \ , \ \ 
\eta_{-}=\Box^{1/2}P_{-}\theta^{2}P_{-} \ \ , \ \ 
\bar{\eta}_{+}=\Box^{1/2}P_{+}\bar{\theta}^{2}P_{+} \ , \label{matrizes}                   
\end{equation}
is the quadratic operator of the free part of the Lagrangian. Equation (\ref{ordem0}) will be the order-$\delta^0$ contribution to the effective potential.

\section{The effective potential using the FAC criterion}

In this section, we shall present the main steps yielding the final expression for the effective potential. The detailed calculation is shown in the Appendix. 

The perturbative effective potential can now be calculated in powers of $\delta$ using the one particle irreducible functions, defined in the expansion (\ref{Gamma Phi barPhi}) of the effective action. We show that, after the optimization procedure, the order-$\delta^{0}$ contribution provides the sum of all one-loop diagrams. In Fig. 1, one can see the diagrammatic sum of the effective potential up to the order $\delta^{1}$ (${\mathcal V}_{eff}^{(1)}$).
%\begin{figure}[ht]
%\begin{center}
%{\includegraphics[bb=115 705 470 750, clip]{Fig1.ps}}  %%nome da Fig. aqui!!!
%\caption{Effective potential up to the order $\delta^1$.}
%\end{center}  
%\label{Fig1}
%\end{figure}
\begin{eqnarray*}
\begin{picture}(350,5) \thicklines
\put(15,0){\circle{25}}\put(30,-3){+}
\put(55,0){\circle{25}}\put(67,0){\line(50,0){20}}\put(73,4){$\phi_0$}\put(91,-3){+}
\put(116,0){\circle{25}}\put(123,-3){$\times$}\put(132,0){$\theta^2$}\put(120,14){$\phi_1$}\put(120,-18){$\phi_1$}\put(145,-3){+}
\put(171,0){\circle{25}}\put(183,0){\line(50,0){20}}\put(189,4){$\phi_1$}\put(208,-3){+}
\put(233,0){\circle{25}}\put(240.5,-3){$\times$}\put(237,14){$\phi_0$}\put(237,-18){$\phi_1$}\put(255,-3){+}
\put(282,0){\circle{25}}\put(289,-3){$\times$}\put(286,14){$\phi_1$}\put(286,-18){$\phi_2$}\put(304,-3){+}
\put(321,-3){$h.c.$}
\end{picture}
\end{eqnarray*}

\vspace{0,1cm}
\begin{center}
Fig. 1: Effective potential up to the order $\delta^1$. 
\end{center}

Note that, by virtue of the $\theta$-dependent propagators, the tadpole diagrams are not identically zero, as is usual in superspace. The first diagram is of order $\delta^{0}$ and corresponds to the first term of (\ref{Gamma Phi barPhi}) for the effective action. Now, we have to fix the mass parameters and write the order-$\delta^0$ effective potential in terms of the optimized parameters, $\mu_0$ and $\bar{\mu}_0$. The FAC criterion is employed as an optimization procedure. In order to calculate the effective potential up to the order zero in $\delta$ (${\mathcal V}_{eff}^{(0)}$), we have to solve, for $\mu_0$ and $\bar{\mu}_0$, the equation
\begin{eqnarray}
c^1(\mu,\bar\mu)&=&0 \ , \label{fac1}
\end{eqnarray}
at $\delta =1$, where $c^1(\mu,\bar\mu)$ corresponds to the $\delta^{1}$ coefficients in the perturbative expansion of the effective potential. In fact, since we split the parameters $M_{ij}$ into a $\theta$-independent ($a_{ij}$) and a $\theta$-dependent ($b_{ij}$) part, and recalling (\ref{choice}), the optimized parameters will be $a_{01}=a$, $b_{11}=b$, and $\rho_{12}=\rho$.  

Using the Feynman rules, one can write the effective potential up to the order $\delta^{1}$ of Fig. 1 as
\begin{eqnarray}
{\mathcal V}_{eff}^{(1)}&=&-\frac{1}{2}\int d^{4}\theta_{1}d^{4}\theta_{2}\delta^{4}(\theta_{1}-\theta_{2})Tr\ln[P^{T}K]\delta^{4}(\theta_{1}-\theta_{2}) \nonumber\\
&&+\delta\int\frac{d^{4}k}{(2\pi)^{4}}K_{1}\left\{2g\bar{b}\int d^{2}\theta\phi_{0}-\left|b\right|^{2}+2gb\int d^{2}\bar{\theta}\bar{\phi}_{0}-\left|b\right|^{2}\right\} \nonumber\\
&&+\delta\int\frac{d^{4}k}{(2\pi)^{4}}K_{2}\left\{-2g\bar{a}\left|b\right|^{2}\int d^{2}\theta\theta^{2}\phi_{1}+\left|a\right|^{2}\left|b\right|^{2}+\rho\bar{M}\left|b\right|^{2}\right. \nonumber\\
&&\left.\hspace{3cm}-2ga\left|b\right|^{2}\int d^{2}\bar{\theta}\bar{\theta}^{2}\bar{\phi}_{1}+\left|a\right|^{2}\left|b\right|^{2}+\bar{\rho}M\left|b\right|^{2}\right\} \ , \label{potencial efetivo}
\end{eqnarray}
where 
\begin{equation}
K_{1}=\frac{1}{\left(k^{2}+\left|M\right|^{2}+\left|a\right|^{2}\right)^{2}\!-\!\left|b\right|^{2}} \ \ , \ \ K_{2}=\frac{1}{\left(k^{2}+\left|M\right|^{2}+\left|a\right|^{2}\right)\left[\left(k^{2}+\left|M\right|^{2}+\left|a\right|^{2}\right)^{2}\!-\!\left|b\right|^{2}\right]} \ . \label{Ks}
\end{equation}

From this equation, we can derive the following simple solution to (\ref{fac1}) at $\delta=1$, before calculating the integrals:
\begin{eqnarray}
a&=&2g\int d^{2}\theta\theta^{2}\phi_{1} \ \ \ , \ \ \ \bar{a}=2g\int d^{2}\bar{\theta}\bar{\theta}^{2}\bar{\phi}_{1} \ ; \nonumber\\
b&=&2g\int d^{2}\theta\phi_{0} \ \ \ , \ \ \ \bar{b}=2g\int d^{2}\bar{\theta}\bar{\phi}_{0} \ ; \nonumber\\
\rho&=&0 \ \ \ , \ \ \ \bar{\rho}=0 \ . \label{abrho}
\end{eqnarray}
This result shows that the optimized parameters $a$ and $b$ are functions of the original couplings and fields of the theory, as we expected.

Recalling that $\phi_{0}$ and $\phi_{1}$ are classical superfields and solving the superspace integrals, we obtain
\begin{eqnarray}
a&=&\bar{a}=2g\langle z_{1}\rangle \ ; \\
b&=&\bar{b}=2g\langle h_{0}\rangle \ ,
\end{eqnarray}
where $\langle z_{1}\rangle$ and $\langle h_{0}\rangle$ are the vacuum expectation values of the scalar and of the scalar auxiliary component fields of $\phi_{1}$ and $\phi_{0}$, respectively.

Replacing this solution into the order-zero contribution (\ref{ordem0}), we obtain
\begin{equation}
{\mathcal V}_{eff}^{(0)}=-\frac{1}{2}tr\int\frac{d^{4}k}{(2\pi)^{4}}\left[\ln\left(1+\frac{\tilde{A}+\tilde{B}}{k^{2}}\right)+\ln\left(1+\frac{\tilde{A}-\tilde{B}}{k^{2}}\right)-2\ln\left(1+\frac{\tilde{A}}{k^{2}}\right)\right] \ , \label{vacuum}
\end{equation} 
where
\begin{equation}
\tilde{A}=\left(\begin{array}{ccc}
          4g^{2}\langle z_{1}\rangle^{2} & 0 & 2mg\langle z_{1}\rangle \\
          0 & m^{2}+4g^{2}\langle z_{1}\rangle^{2} & 0 \\
          2mg\langle z_{1}\rangle & 0 & m^{2}
          \end{array}\right) \ \ \ , \ \ \ 
          \tilde{B}=\left(\begin{array}{ccc}
                 0 & 0 & 0 \\
                 0 & 2g\langle h_{0}\rangle & 0 \\
                 0 & 0 & 0
                \end{array}\right) \ .
\end{equation}
 
After regularization and renormalization procedure, the result reads as below:
\begin{eqnarray}
{\mathcal V}_{eff}^{(0)}=-\frac{1}{(8\pi)^{2}}\left\{(m^{2}+4g^{2}\langle z_{1}\rangle^{2})^{2}\ln\left[1-\frac{4g^{2}\langle h_{0}\rangle^{2}}{(m^{2}+4g^{2}\langle z_{1}\rangle^{2})^{2}}\right]\right. \hspace{1,5cm}\nonumber\\
\left.+4g\langle h_{0}\rangle(m^{2}+4g^{2}\langle z_{1}\rangle^{2})\ln\frac{m^{2}+4g^{2}\langle z_{1}\rangle^{2}+2g\langle h_{0}\rangle}{m^{2}+4g^{2}\langle z_{1}\rangle^{2}-2g\langle h_{0}\rangle}\right. \nonumber\\
\left.+4g^{2}\langle h_{0}\rangle^{2}\ln\left[(m^{2}+4g^{2}\langle z_{1}\rangle^{2})^{2}-4g^{2}\langle h_{0}\rangle^{2}\right]\right\} \ . \hspace{0,9cm}  
\end{eqnarray}

Now, using the notation of \cite{Helayel} and writing $\langle z_{1}\rangle=\frac{1}{\sqrt{2}}A_{1}$ and $\langle h_{0}\rangle=\frac{1}{\sqrt{2}}F_{0}$, we obtain
\begin{eqnarray}
{\mathcal V}_{eff}^{(0)}=-\frac{1}{(8\pi)^{2}}\left\{(m^{2}+2g^{2}A_{1}^{2})^{2}\ln\left[1-\frac{2g^{2}F_{0}^{2}}{(m^{2}+2g^{2}A_{1}^{2})^{2}}\right]\right. \hspace{1,8cm}\nonumber\\
\left.+2\sqrt{2}gF_{0}(m^{2}+2g^{2}A_{1}^{2})\ln\frac{m^{2}+2g^{2}A_{1}^{2}+\sqrt{2}gF_{0}}{m^{2}+2g^{2}A_{1}^{2}-\sqrt{2}gF_{0}}\right. \nonumber\\
\left.+2g^{2}F_{0}^{2}\ln\left[(m^{2}+2g^{2}A_{1}^{2})^{2}-2g^{2}F_{0}^{2}\right]\right\} \ . \hspace{1,6cm} \label{final} 
\end{eqnarray}
This is the Coleman-Weinberg potential for the O'Raifeartaigh model and it represents the sum of all one-loop supergraphs.

\section{Optimized solutions using the PMS criterion}

In this section we are going to derive the solutions to the optimized parameters using the PMS criterion. 

According to the PMS we need to solve the following equations:
\begin{equation}
\left.\frac{\partial {\mathcal V}_{eff}^{(1)}(a,b,\rho)}{\partial a}\right|_{a=a_0,\delta=1}=0 \ \ , \ \ \left.\frac{\partial {\mathcal V}_{eff}^{(1)}(a,b,\rho)}{\partial b}\right|_{b=b_0,\delta=1}=0 \ \ , \ \ 
\left.\frac{\partial {\mathcal V}_{eff}^{(1)}(a,b,\rho)}{\partial\rho}\right|_{\rho=\rho_0,\delta=1}=0 \ , \label{eqPMS}
\end{equation} 
for the $a$, $b$, and $\rho$ parameters. In order to find the solutions to the above equations, we rewrite (\ref{vacuum}) for the vacuum diagram as

\begin{equation}
{\mathcal V}_{eff}^{(0)}=-\frac{1}{2}\int\frac{d^{4}k}{(2\pi)^{4}}\ln\left[1-\frac{\left|b\right|^{2}}{\left(k^{2}+\left|M\right|^{2}+\left|a\right|^{2}\right)^{2}}\right] \ . \label{ln}
\end{equation}

The PMS equation for the $a$ parameter is given by
\begin{eqnarray}
\frac{\partial {\mathcal V}_{eff}^{(1)}}{\partial a}=&&\int\frac{d^{4}k}{(2\pi)^{4}}K_{2}\left|b\right|^{2}\left\{-\bar{a}+2g\int d^{2}\bar{\theta}\bar{\theta}^{2}\bar{\phi}_{1}\right\} \nonumber\\
&+&\int\frac{d^{4}k}{(2\pi)^{4}}2\bar{a}K_{1}^{2}(k^{2}+\left|M\right|^{2}+\left|a\right|^{2})\left\{\bar{b}\left(b-2g\int d^{2}\theta\phi_{0}\right)+b\left(\bar{b}-2g\int d^{2}\bar{\theta}\bar{\phi}_{0}\right)\right\} \nonumber\\
&+&\!\!\int\!\!\frac{d^{4}k}{(2\pi)^{4}}\bar{a}\left|b\right|^{4}\!K_{2}^{2}\!\left\{\!\bar{a}\!\left(a-2g\!\!\int \!\!d^{2}\theta\theta^{2}\phi_{1}\right)\!\!+\!a\!\left(\bar{a}-2g\!\!\int \!\!d^{2}\bar{\theta}\bar{\theta}^{2}\bar{\phi}_{1}\right)\!+\!\rho\bar{M}\!+\!\bar{\rho}M\!\right\} \nonumber\\
&+&\!\!\int\!\!\frac{d^{4}k}{(2\pi)^{4}}3\bar{a}\left|b\right|^{2}\!K_{1}^{2}\!\left\{\!\bar{a}\!\left(2g\!\!\int \!\!d^{2}\theta\theta^{2}\phi_{1}-a\right)\!\!+\!a\!\left(2g\!\!\int \!\!d^{2}\bar{\theta}\bar{\theta}^{2}\bar{\phi}_{1}-\bar{a}\right)\!-\!\rho\bar{M}\!-\!\bar{\rho}M\!\right\} \nonumber\\
&=&0 \ , \label{solPMS}
\end{eqnarray}
where $K_{1}$ and $K_{2}$ are defined by (\ref{Ks}) and we have an analogous equation for $b$ and $\rho$. From this we see that the same result as (\ref{abrho}) is obtained, and when we plug these solutions into (\ref{potencial efetivo}) we obtain the very same solution as (\ref{final}).
 
Here, it should be emphasized that (\ref{abrho}) is not the unique solution of (\ref{fac1}) and (\ref{eqPMS}). If we regularize and renormalize before the optimization procedure we find other solutions. However, these are not physical, as can be checked by plugging these solutions into the effective potential. 

Also, there is no guarantee that the PMS and FAC criteria give always the same result to all orders of perturbation theory.

\section{Concluding Remarks}
We have applied superspace and supergraph techniques to carry out the LDE for the O'Raifeartaigh \cite{O'Raifeartaigh} model, which is the minimal way to realize spontaneous breaking of SUSY in the matter sector. The method explicitly breaks SUSY because the arbitrary mass parameter is a superfield and, when expanded, yields soft breaking terms in the Lagrangian.
We have shown that, from a perturbative calculation with new propagators and interactions, it was possible to reproduce the sum of infinite diagrams of a specific set. In particular, it has been shown that, after an optimization procedure to order $\delta^{1}$, which consists of taking into account just a few diagrams, the order-$\delta^{0}$ effective potential reproduces the sum of all one-loop diagrams. This is a very suggestive result, because if we now calculate the effective potential up to the order $\delta^2$, we will find the type of two-loop diagrams shown in Fig. 2.
%\begin{figure}[ht]
%\begin{center}
%{\includegraphics[bb=115 715 470 750, clip]{Fig2.ps}}  %%nome da Fig. aqui!!!
%\caption{Order-$\delta^2$ two-loop diagram.}
%\end{center}  
%\label{Fig2}
%\end{figure}
\begin{eqnarray*}
\begin{picture}(350,5) \thicklines
\put(170,0){\circle{25}}
\put(170,-12){\line(0,5){24}}
\end{picture}
\end{eqnarray*}

\vspace{0,1cm}
\begin{center}
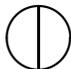
Fig. 2: Order-$\delta^2$ two-loop diagram. 
\end{center} 

Now, after the optimization procedure, we shall get nonperturbative corrections which include a set of two-loop diagrams. In this case, if we use the FAC criterion, it will be necessary to evaluate (\ref{fac}) to order $\delta^{3}$. On the other hand, if we use the PMS criterion, we need just to evaluate (\ref{PMS}) up to the order $\delta^{2}$, which looks easier. In both cases, we expect that we shall not be able to find a simple analytic solution, as was found here. However, such an endeavor is very important, in order to study in a systematic way the effects of nonperturbative corrections to SUSY and \textit{R} symmetry breaking. This is a work in progress and we shall report on it elsewhere \cite{nos2}.

Also, the efficacy of the method whenever applied to superspace and superfields should be tested for the SUSY breaking realized \`a la Fayet-Iliopoulos \cite{Fayet-Iliopoulos}. In this particular situation, the results of the work of Ref. \cite{Helayel 2} show that the superspace calculations are much more involved than in the case of the O'Raifeartaigh's realization; nontrivial mixings between the matter and gauge potential superfields show up that yield a whole discussion on the $R_\xi$-type gauge fixing in superspace. So, the Fayet-Iliopoulos model for SUSY breaking sets up an appropriate scenario for testing the LDE in connection with superfield techniques and the treatment of the matter-gauge sector mixings shall require due care which may compell us to better understand a number of technicalities inherent to the LDE.

\section{Acknowledgements}

M. C. B. Abdalla and J. A. Helay\"{e}l-Neto acknowledge CNPq for support. Carlos R. Senise Jr. thanks CAPES-Brazil for financial support.

\section*{Appendix: Explicit calculation of the effective potential in superspace}

In this Appendix, we present a detailed calculation of the order-$\delta^{0}$ effective potential in superspace using superspace techniques, following the same approach as in \cite{Nibbelink}.

The quadratic part of the free Lagrangian is given by 
\begin{equation}
{\mathcal L}_{0}^{\delta}=\int d^{4}\theta\bar{\phi}_{i}\phi_{i}-\left[\int d^{2}\theta\left(M\phi_{1}\phi_{2}+a\phi_{0}\phi_{1}+\frac{1}{2}b\theta^{2}\phi_{1}^{2}\right)+h.c.\right] \ . \label{ap1}  
\end{equation}

We can rewrite this equation using matrix notation, as
\begin{equation}
{\mathcal L}_{0}^{\delta}=\int d^{4}\theta\bar{\Phi}\Phi-\frac{1}{2}\int d^{2}\theta\Phi^{T}(A+B\theta^{2})\Phi-\frac{1}{2}\int d^{2}\bar{\theta}\bar{\Phi}(\bar{A}+\bar{B}\bar{\theta}^{2})\bar{\Phi}^{T} \ , \label{ap2}
\end{equation}
where $A$ and $B$ are given by (\ref{matrizes}) and 
\begin{equation}
\Phi=\left(\begin{array}{c}
          \phi_0 \\
          \phi_1 \\
          \phi_2
          \end{array}\right) \ \ , \ \ 
\bar{\Phi}=\left(\begin{array}{ccc}
          \bar{\phi}_0 & \bar{\phi}_1 & \bar{\phi}_2 
          \end{array}\right) \ \ . \label{ap3}                   
\end{equation}

To work directly in superspace, using the spurions $\theta^2$ and $\bar{\theta}^2$, it is now convenient to present some useful operators, defined in Ref. \cite{Nibbelink}:
\begin{equation}
\eta_{\pm}=\Box^{1/2}P_{\pm}\theta^{2}P_{\pm} \ \ \ , \ \ \ \bar{\eta}_{\pm}=\Box^{1/2}P_{\pm}\bar{\theta}^{2}P_{\pm} \ . \label {ap5}
\end{equation}

It is important to note that the $\eta_{+}\bar{\eta}_{+}$ and $\bar{\eta}_{+}\eta_{+}$ products, and similarly $\eta_{-}\bar{\eta}_{-}$ and $\bar{\eta}_{-}\eta_{-}$, are not equal, as can be seen by
\begin{eqnarray}
\eta_{+}\bar{\eta}_{+}=\frac{1}{16}\bar{D}^{2}\theta^{2}\bar{\theta}^{2}D^{2} \ \ \ &,& \ \ \ \bar{\eta}_{+}\eta_{+}=\Box P_{+}\theta^{2}\bar{\theta}^{2}P_{+} \ ; \nonumber\\
\eta_{-}\bar{\eta}_{-}=\Box P_{-}\theta^{2}\bar{\theta}^{2}P_{-} \ \ \ &,& \ \ \ \bar{\eta}_{-}\eta_{-}=\frac{1}{16}D^{2}\theta^{2}\bar{\theta}^{2}\bar{D}^{2} \ . \label{ap6} 
\end{eqnarray}

The spurion operators $\eta_{+}$ and $\bar{\eta}_{+}$ have interesting algebraic properties. They generate a Clifford algebra: $\left\{\eta_{+},\bar{\eta}_{+}\right\}=\textbf{1}_{+}$, where the combination $\textbf{1}_{+}=\eta_{+}\bar{\eta}_{+}+\bar{\eta}_{+}\eta_{+}$ plays the role of the identity, since we have $\eta_{+}\textbf{1}_{+}=\textbf{1}_{+}\eta_{+}=\eta_{+}$ and $\bar{\eta}_{+}\textbf{1}_{+}=\textbf{1}_{+}\bar{\eta}_{+}=\bar{\eta}_{+}$. Therefore, we can interpret the combination
\begin{equation}
A_{+}=A_{11}\bar{\eta}_{+}\eta_{+}+A_{12}\bar{\eta}_{+}+A_{21}\eta_{+}+A_{22}\eta_{+}\bar{\eta}_{+} \label{ap7}
\end{equation}
as a $2\times 2$ matrix:
\begin{equation}
A_{+}=\left(\begin{array}{cc}
                   A_{11} & A_{12} \\
                   A_{21} & A_{22}
                   \end{array}
                   \right)_{\eta_{+}} \ . \label{ap8}
\end{equation}

We can also define the projection operator
\begin{equation}
\bot_{+}=P_{+}-\textbf{1}_{+}=P_{+}-\eta_{+}\bar{\eta}_{+}-\bar{\eta}_{+}\eta_{+} \ , \label{ap9}
\end{equation}
which is perpendicular to any $A_{+}$, i.e., $\bot_{+}A_{+}=A_{+}\bot_{+}=0$. 

Finally, we define the trace of $A_{+}$ as
\begin{eqnarray}
trA_{+}&=&\left(\begin{array}{cc}
                       trA_{11} & trA_{12} \\
                       trA_{21} & trA_{22}
                       \end{array}
                       \right)_{\eta_{+}} \nonumber\\
&=&trA_{11}\bar{\eta}_{+}\eta_{+}+trA_{12}\bar{\eta}_{+}+trA_{21}\eta_{+}+trA_{22}\eta_{+}\bar{\eta}_{+} \ . \label{ap10}
\end{eqnarray} 

Once $\eta_{-}$ and $\bar{\eta}_{-}$ satisfy the same algebra as $\eta_{+}$ and $\bar{\eta}_{+}$, equations similar to (\ref{ap7} - \ref{ap10}) can be written for $A_{-}$, $\bot_{-}$, and $trA_{-}$, replacing $\eta_{+}$ and $\bar{\eta}_{+}$ by $\eta_{-}$ and $\bar{\eta}_{-}$.

Using the operators $P_{\pm}$, $\eta_{\pm}$, and $\bar{\eta}_{\pm}$, the free part of the action (\ref{ap2}) can be writen as
\begin{equation}
{\mathcal L}_{0}^{\delta}=\int d^{4}\theta\left\{\bar{\Phi}\Phi+\frac{1}{2}\Phi^T\left(AP_{-}+B\frac{1}{\Box^{1/2}}\eta_{-}\right)\frac{D^{2}}{4\Box}\Phi+\frac{1}{2}\bar{\Phi}\left(\bar{A}P_{+}+\bar{B}\frac{1}{\Box^{1/2}}\bar{\eta}_{+}\right)\frac{\bar{D}^{2}}{4\Box}\bar{\Phi}^T\right\} \ , \label{ap11}
\end{equation}
and the expression for the vacuum diagram is given by 
\begin{equation}
{\mathcal V}_{eff}^{(0)}=-\frac{1}{2}\int d^{4}\theta_{12}\delta_{21}Tr\ln[P^{T}K]\delta_{21} \ , \label{ap12}
\end{equation}
where $K$ is the quadratic operator of the free part of the Lagrangian and $P$ is the matrix defined by (\ref{P}). To evaluate the vacuum diagram, we write the Lagrangian ${\mathcal L}_{0}^{\delta}$ as
\begin{equation}
{\mathcal L}_{0}^{\delta}=\frac{1}{2}\int d^{4}\theta\left(\begin{array}{cc}
                                     \Phi^T & \bar{\Phi}
                                     \end{array}
                                     \right)K\left(\begin{array}{c}
                                                   \Phi \\
                                                   \bar{\Phi}^T
                                                   \end{array}
                                                   \right)
=\frac{1}{2}\int d^{4}\theta\left(\begin{array}{cc}
                                     \Phi^T & \bar{\Phi}
                                     \end{array}
                                     \right)\left(\begin{array}{cc}
                                                  K_{\Phi\Phi} & K_{\Phi\bar{\Phi}} \\
                                                  K_{\bar{\Phi}\Phi} & K_{\bar\Phi\bar\Phi}
                                                  \end{array}
                                                  \right)\left(\begin{array}{c}
                                                   \Phi \\
                                                   \bar{\Phi}^T
                                                   \end{array}
                                                   \right) \ , \label{ap14}                                                   
\end{equation}
where $K$ is defined by (\ref{K}).

From Eqs.(\ref{P}) and (\ref{K}) we write
\begin{equation}
P^{T}K=\left(\begin{array}{cc}
       P_{+} & K_{\bar\Phi\bar\Phi} \\
       K_{\Phi\Phi} & P_{-}
       \end{array}
       \right)
=\left( \begin{array}{cc}
P_{+} & C \\
\bar{C} & P_{-}
\end{array}
\right) \ , \label{ap16}
\end{equation}
with $C$ and $\bar{C}$ given by (\ref{K}).

Now, writing $P^TK$ as
\begin{eqnarray}
\left( \begin{array}{cc}
P_{+} & 0 \\
0 & P_{-}
\end{array}
\right)\left( \begin{array}{cc}
1 & C \\
\bar{C} & 1
\end{array}
\right) \ , \nonumber 
\end{eqnarray}
the expression for the vacuum diagram (\ref{ap12}) reads
\begin{equation}
{\mathcal V}_{eff}^{(0)}=-\frac{1}{2}\int d^{4}\theta_{12}\delta_{21}\{tr\ln P_{+}+tr\ln P_{-}+Tr\ln[1+Z]\}\delta_{21} \ , \label{ap17}
\end{equation} 
where 
\begin{equation}
Z=\left( \begin{array}{cc}
0 & C \\
\bar{C} & 0
\end{array}
\right) \ . \label{ap18}
\end{equation}
From (\ref{ap17}), ${\mathcal V}_{eff}^{0}$ splits into three parts: a $P_{+}$ contribution, a $P_{-}$ contribution, and a contribution proportional to $C$ and $\bar{C}$. 

The $P_{+}$ contribution is given by
\begin{equation}
{\mathcal V}_{eff}^{(0)P_{+}}=-\frac{1}{2}\int d^{4}\theta_{12}\delta_{21}tr\ln P_{+}\delta_{21} \ . \label{ap19}
\end{equation}
Since $\delta^4(\theta_2-\theta_1)P_{+}\delta^4(\theta_2-\theta_1)=-\delta^4(\theta_2-\theta_1)\frac{1}{p^2}$, we obtain ${\mathcal V}_{eff}^{(0)P_{+}}=0$ and, identically, ${\mathcal V}_{eff}^{(0)P_{-}}=0$. The contribution from the last term in (\ref{ap17}) is given by
\begin{equation}
{\mathcal V}_{eff}^{(0)C}=-\frac{1}{2}\int d^{4}\theta_{12}\delta_{21}Tr\ln(1+Z)\delta_{21} \ . \label{ap20}
\end{equation}
We introduce a continuous parameter $0\leq\lambda\leq 1$ in front of the off-diagonal terms ($C$ and $\bar{C}$), so that
\begin{equation}
{\mathcal V}_{eff}^{(0)C\lambda}=-\frac{1}{2}\int d^{4}\theta_{12}\delta_{21}Tr\ln(1+\lambda Z)\delta_{21} \ , \label{ap21}
\end{equation}
and
\begin{equation}
{\mathcal V}_{eff}^{(0)C}=\int_{0}^{1}d\lambda\frac{d}{d\lambda}{\mathcal V}_{eff}^{(0)C\lambda} \ . \label{ap22}
\end{equation}    

Differentiating (\ref{ap21}) with respect to $\lambda$ we obtain
\begin{equation}
\frac{d}{d\lambda}{\mathcal V}_{eff}^{(0)C\lambda}=-\frac{1}{2}\int d^{4}\theta_{12}\delta_{21}Tr[Z-\lambda Z^{2}+\lambda^{2}Z^{3}-\lambda^{3}Z^{4}+...]\delta_{21} \ . \label{ap23}
\end{equation}

Since
\begin{equation}
Z^2=\left( \begin{array}{cc}
C\bar{C} & 0 \\
0 & \bar{C}C
\end{array}
\right) \label{ap24}
\end{equation}
is block diagonal, it follows that odd powers of $Z$ are necessarily off-diagonal in the real basis of the chiral multiplets. The trace $Tr$ in this basis is the sum of the traces in the complex basis of the block diagonal parts; hence the odd powers of $Z$ do not contribute, and (\ref{ap23}) reads
\begin{eqnarray}
\frac{d}{d\lambda}{\mathcal V}_{eff}^{(0)C\lambda}=-\frac{1}{2}\int d^{4}\theta_{12}\delta_{21}Tr[-\lambda Z^{2}-\lambda^{3}Z^{4}-\lambda^{5}Z^{6}-...]\delta_{21} \ . \label{ap25}
\end{eqnarray}

Integrating with respect to $\lambda$ we obtain
\begin{equation}
{\mathcal V}_{eff}^{(0)C}=-\frac{1}{4}\int d^{4}\theta_{12}\delta_{21}tr[\ln(P_{+}-C\bar{C})+\ln(P_{-}-\bar{C}C)]\delta_{21} \ . \label{ap26}
\end{equation}

Writing
\begin{eqnarray}
C\bar{C}&=&\frac{\bar{A}A}{\Box}P_{+}+\frac{\bar{A}B}{\Box^{3/2}}\eta_{+}+\frac{\bar{B}A}{\Box^{3/2}}\bar{\eta}_{+}+\frac{\bar{B}B}{\Box^2}\bar{\eta}_{+}\eta_{+} \ , \nonumber\\
\bar{C}C&=&\frac{A\bar{A}}{\Box}P_{-}+\frac{B\bar{A}}{\Box^{3/2}}\eta_{-}+\frac{A\bar{B}}{\Box^{3/2}}\bar{\eta}_{-}+\frac{B\bar{B}}{\Box^2}\eta_{-}\bar{\eta}_{-} \ , \label {ap27}
\end{eqnarray}
and using (\ref{ap8}), we derive the following relations:
\begin{eqnarray}
P_{-}-\bar{C}C&=&\bot_{-}\left(1-\frac{A\bar{A}}{\Box}\right)+\textbf{1}_{-}-E_{-} \ , \nonumber\\
P_{+}-C\bar{C}&=&\bot_{+}\left(1-\frac{\bar{A}A}{\Box}\right)+\textbf{1}_{+}-E_{+} \ , \label{ap28}
\end{eqnarray}
where
\begin{eqnarray}
E_{-}&=&\left(\begin{array}{cc}
            \frac{A\bar{A}}{\Box} & \frac{A\bar{B}}{\Box^{3/2}} \\
            \frac{B\bar{A}}{\Box^{3/2}} & \frac{1}{\Box}\left(A\bar{A}+\frac{B\bar{B}}{\Box}\right)
            \end{array}
            \right)_{\eta_{-}}
=\left(\begin{array}{cc}
            \alpha & \beta \\
            \gamma & \delta
            \end{array}
            \right)_{\eta_{-}} \ , \nonumber\\
E_{+}&=&\left(\begin{array}{cc}
              \frac{1}{\Box}\left(\bar{A}A+\frac{\bar{B}B}{\Box}\right) & \frac{\bar{B}A}{\Box^{3/2}} \\
              \frac{\bar{A}B}{\Box^{3/2}} & \frac{\bar{A}A}{\Box}
              \end{array}
              \right)_{\eta_{+}}
=\left(\begin{array}{cc}
            \delta^T & \beta^T \\
            \gamma^T & \alpha^T
            \end{array}
            \right)_{\eta_{+}} \ . \label{ap29}
\end{eqnarray}  

With these relations, (\ref{ap26}) reads
\begin{eqnarray}
{\mathcal V}_{eff}^{(0)C}=-\frac{1}{4}\int d^{4}\theta_{12}\delta_{21}tr\left\{\ln\left[\bot_{+}\left(1-\frac{A\bar{A}}{\Box}\right)+\textbf{1}_{+}-E_{+}\right]+\right. \hspace{1cm} \nonumber\\
\left.+\ln\left[\bot_{-}\left(1-\frac{A\bar{A}}{\Box}\right)+\textbf{1}_{-}-E_{-}\right]\right\}\delta_{21} \ . \label{ap30}
\end{eqnarray}

Equation (\ref{ap30}) splits into two parts: a $\bot_{\pm}$ contribution and another one proportional to $(\textbf{1}-E)_{\pm}$.

First we compute the $(\textbf{1}_{-}-E_{-})$ contribution:
\begin{equation}
{\mathcal V}_{eff}^{(0)\eta_{-}}=-\frac{1}{4}\int d^{4}\theta_{12}\delta_{21}tr[\ln(\textbf{1}_{-}-E_{-})]_{2}\delta_{21} \ . \label{ap31}
\end{equation}

We again introduce a continuous parameter $0\leq\lambda\leq 1$:
\begin{equation}
{\mathcal V}_{eff}^{(0)\eta_{-}\lambda}=-\frac{1}{4}\int d^{4}\theta_{12}\delta_{21}tr[\ln(\textbf{1}_{-}-\lambda E_{-})]_{2}\delta_{21} \ . \label{ap32}
\end{equation}

Differentiating with respect to $\lambda$,
\begin{equation}
\frac{d}{d\lambda}{\mathcal V}_{eff}^{(0)\eta_{-}\lambda}=\frac{1}{4}\int d^{4}\theta_{12}\delta_{21}trF[E_{-}]_{2}\delta_{21} \ \ \ \ \ \ \ \ , F[E]=E(\textbf{1}-\lambda E)^{-1} \ . \label{ap33} 
\end{equation}
Explicitly, this gives
\begin{eqnarray}
\frac{d}{d\lambda}{\mathcal V}_{eff}^{(0)\eta_{-}\lambda}=\frac{1}{4}\int d^{4}\theta_{12}\delta_{21}tr\left\{F_{11}[E_{-}]\bar{\eta}_{-}\eta_{-}+F_{12}[E_{-}]\bar{\eta}_{-}+\right. \hspace{1,2cm} \nonumber\\
\left.+F_{21}[E_{-}]\eta_{-}+F_{22}[E_{-}]\eta_{-}\bar{\eta}_{-}\right\}_{2}\delta_{21} \ . \label{ap34}
\end{eqnarray}

Using the relations 
\begin{equation}
\int d^{4}\theta_{12}\delta_{21}[F\eta_{\pm}\bar{\eta}_{\pm}]_{2}\delta_{21}=\int d^{4}\theta_{12}\delta_{21}[F\bar{\eta}_{\pm}\eta_{\pm}]_{2}\delta_{21}=\int d^{4}\theta\theta^{2}\bar{\theta}^{2}F_{2} \  \label{ap35}
\end{equation}
and
\begin{eqnarray}
\int d^{4}\theta_{12}\delta_{21}[F\eta_{\pm}]_{2}\delta_{21}&=&\int d^{4}\theta\theta^{2}\left[F\frac{1}{\Box^{1/2}}\right]_{2} \ ; \nonumber\\
\int d^{4}\theta_{12}\delta_{21}[F\bar{\eta}_{\pm}]_{2}\delta_{21}&=&\int d^{4}\theta\bar{\theta}^{2}\left[F\frac{1}{\Box^{1/2}}\right]_{2} \ , \label{ap36}
\end{eqnarray}
we rewrite (\ref{ap34}) as
\begin{equation}
\frac{d}{d\lambda}{\mathcal V}_{eff}^{(0)\eta_{-}\lambda}=\frac{1}{4}\int d^{4}\theta tr\left\{(F_{11}[E_{-}]+F_{22}[E_{-}])\theta^{2}\bar{\theta}^{2}+\frac{1}{\Box^{1/2}}F_{21}[E_{-}]\theta^{2}+\frac{1}{\Box^{1/2}}F_{12}[E_{-}]\bar{\theta}^{2}\right\} \ . \label{ap37}
\end{equation}
The very same result is achieved for the $(\textbf{1}_{+}-E_{+})$ contribution. The sum reads
\begin{eqnarray}
\frac{d}{d\lambda}{\mathcal V}_{eff}^{(0)\eta\lambda}\!&=&\!\frac{d}{d\lambda}{\mathcal V}_{eff}^{(0)\eta_{-}\lambda}+\frac{d}{d\lambda}{\mathcal V}_{eff}^{(0)\eta_{+}\lambda} \nonumber\\
&=&\!\frac{1}{2}\int\!d^{4}\theta tr\!\left\{\!(F_{11}[E_{-}]+F_{22}[E_{-}])\theta^{2}\bar{\theta}^{2}\!+\!\frac{1}{\Box^{1/2}}F_{21}[E_{-}]\theta^{2}\!+\!\frac{1}{\Box^{1/2}}F_{12}[E_{-}]\bar{\theta}^{2}\right\} \ . \label{ap38}
\end{eqnarray}

In our case, due to the no-dependence of the $A$, $B$ matrices with the $\theta$'s, we have ${\mathcal V}_{eff}^{(0)\eta\theta^{2}}={\mathcal V}_{eff}^{(0)\eta\bar{\theta}^{2}}=0$, and the only remaining term is the $\theta^{2}\bar{\theta}^{2}$ contribution.

Using (\ref{ap29}), we write
\begin{equation}
F_{11}[E_{-}]+F_{22}[E_{-}]=-\frac{d}{d\lambda}\left[\ln(1-\lambda\alpha)+\ln\left(1-\lambda\delta-\frac{\lambda^{2}\beta\gamma}{1-\lambda\alpha}\right)\right] \ , \label{ap39}
\end{equation}
and the $\theta^2\bar{\theta}^2$ contribution is written as
\begin{eqnarray}
\frac{d}{d\lambda}{\mathcal V}_{eff}^{(0)\eta\lambda\theta^2\bar{\theta}^2}&=&\frac{1}{2}\int d^{4}\theta tr\left\{(F_{11}[E_{-}]+F_{22}[E_{-}])\theta^{2}\bar{\theta}^{2}\right\} \nonumber\\
&=&-\frac{1}{2}\int d^{4}\theta\frac{d}{d\lambda}tr\left[\ln(1-\lambda\alpha)+\ln\left(1-\lambda\delta-\frac{\lambda^{2}\beta\gamma}{1-\lambda\alpha}\right)\right]\theta^{2}\bar{\theta}^{2} \ . \label{ap40}
\end{eqnarray}
Integrating with respect to $\lambda$,
\begin{eqnarray}
{\mathcal V}_{eff}^{(0)\eta\theta^2\bar{\theta}^2}&=&-\frac{1}{2}\int d^{4}\theta tr\left[\ln(1-\alpha)+\ln\left(1-\delta-\frac{\beta\gamma}{1-\alpha}\right)\right]\theta^{2}\bar{\theta}^{2} \nonumber\\ 
&=&-\int d^{4}\theta\theta^2\bar{\theta}^2\left\{trL_0(A\bar{A})+\frac{1}{2}K(A\bar{A},B)\right\} \ , \label{ap43}
\end{eqnarray}
where we used the relation
\begin{equation}
1-\delta-\frac{\beta\gamma}{1-\alpha}=\frac{1}{\Box}\left\{\Box-A\bar{A}-\frac{B\bar{B}}{\Box-A\bar{A}}\right\} \ , \label{ap42}
\end{equation}
and
\begin{equation}
L_0(A\bar{A})=\int\frac{d^{4}p}{(2\pi)^{4}}\ln\left(1+\frac{A\bar{A}}{p^2}\right) \ \ \ , \ \ \ K(A\bar{A},B)=\int\frac{d^{4}p}{(2\pi)^{4}}tr\ln\left(1-\frac{B\bar{B}}{(p^2+A\bar{A})^2}\right) \ . \label{ap44}
\end{equation}

The last contribution comes from the terms proportional to $\bot_{\pm}$ in (\ref{ap30}). The $\bot_{-}$ contribution is given by
\begin{eqnarray}
{\mathcal V}_{eff}^{(0)\bot_{-}}&=&-\frac{1}{4}\int d^{4}\theta_{12}\delta_{21}tr\ln\left[\bot_{-}\left(1-\frac{A\bar{A}}{\Box}\right)\right]\delta_{21} \nonumber\\
&=&-\frac{1}{4}\int d^{4}\theta_{12}\delta_{21}tr\ln\left[(P_{-}-\bar{\eta}_{-}\eta_{-}-\eta_{-}\bar{\eta}_{-})\left(1+\frac{A\bar{A}}{p^2}\right)\right]\delta_{21} \ . \label{ap54}
\end{eqnarray}
Using that $\delta^4(\theta_2-\theta_1)P_{+}\delta^4(\theta_2-\theta_1)=-\delta^4(\theta_2-\theta_1)\frac{1}{p^2}$ and the relation (\ref{ap35}), we obtain
\begin{equation}
{\mathcal V}_{eff}^{(0)\bot_{-}}=\int d^{4}\theta tr\left[\frac{1}{4}L_{1}(A\bar{A})+\frac{1}{2}L_{0}(A\bar{A})\theta^{2}\bar{\theta}^{2}\right] \ , \label{ap55}
\end{equation}
where $L_0(A\bar{A})$ is given by (\ref{ap44}) and
\begin{equation}
L_1(A\bar{A})=\int\frac{d^{4}p}{(2\pi)^{4}}\frac{1}{p^2}\ln\left(1+\frac{A\bar{A}}{p^2}\right) \ . \label{ap56}
\end{equation}

The $\bot_{+}$ contribution is exactly the same and we have
\begin{equation}
{\mathcal V}_{eff}^{(0)\bot}={\mathcal V}_{eff}^{(0)\bot_{-}}+{\mathcal V}_{eff}^{(0)\bot_{+}}=\int d^{4}\theta tr\left[\frac{1}{2}L_{1}(A\bar{A})+L_{0}(A\bar{A})\theta^{2}\bar{\theta}^{2}\right] \ , \label{ap57}
\end{equation}
which is the total contribution of the projection operators $\bot_{\pm}$.

Using Eqs. (\ref{ap43}) and (\ref{ap57}), the expression for the vacuum diagram, which is exactly the same as (\ref{vacuum}), is given by
\begin{eqnarray}
{\mathcal V}_{eff}^{(0)}&=&{\mathcal V}_{eff}^{(0)\eta\theta^{2}\bar{\theta}^{2}}+{\mathcal V}_{eff}^{(0)\bot} \nonumber\\
&=&-\frac{1}{2}K(A\bar{A},B) \nonumber\\
&=&-\frac{1}{2}tr\left[L_0(\tilde{A}+\tilde{B})+L_0(\tilde{A}-\tilde{B})-2L_0(\tilde{A})\right] \ , \label{ap59}
\end{eqnarray}
where $\tilde{A}=A\bar{A}$ and $\tilde{B}=(B\bar{B})^{1/2}$.


\begin{thebibliography}{99}

\bibitem{NS} 
  A.~E.~Nelson and N.~Seiberg,
  \textit{R symmetry breaking versus SUSY breaking},
  Nucl.\ Phys.\ B {\bf416} (1994) 46 [hep-ph/9309229].
  
\bibitem{IS}
  K.~Intriligator and N.~Seiberg,
  \textit{Lectures on SUSY breaking},
  Class.\ Quant.\ Grav.\ {\bf24} (2007) S741 [hep-ph/0702069].
  
\bibitem{O'Raifeartaigh}
  L.~O'Raifeartaigh,
  \textit{Spontaneous symmetry breaking for chiral scalar superfields},
  Nucl.\ Phys.\ B {\bf 96} (1975) 331.

\bibitem{ISS1}
  K.~Intriligator, N.~Seiberg and D.~Shih,
  \textit{Dynamical SUSY breaking in meta-stable vacua},
  JHEP {\bf04} (2006) 021 [hep-th/0602239].
  
\bibitem{ISS2}
  K.~Intriligator, N.~Seiberg and D.~Shih,
  \textit{SUSY breaking, R-symmetry breaking and metastable vacua},
  JHEP {\bf07} (2007) 017 [hep-th/0703281].

\bibitem{Shih}
  D.~Shih,
  \textit{Spontaneous R-symmetry breaking in O'Raifeartaigh models},
  JHEP {\bf02} (2008) 091 [hep-th/0703196].
  
\bibitem{Marques}
  L.~G.~Aldrovandi and D.~Marqu\'es,
  \textit{SUSY and R-symmetry breaking in models with non-canonical K\"{a}hler potential},
  JHEP {\bf05} (2008) 022, arXiv:0803.4163 [hep-th]. 

\bibitem{CW} 
  S.~Coleman and E.~Weinberg, 
  \textit{Radiactive corrections as the origin of spontaneous symmetry breaking},
  Phys.\ Rev.\ D {\bf 7} (1973) 1888.
  
\bibitem{Jac}
  R.~Jackiw,
  \textit{Functional evaluation of the effective potential},
  Phys.\ Rev.\  D {\bf 9}, 1686 (1974).
  
\bibitem{delta1}
  A.~Okopinska,
  \textit{Nonstandard expansion techniques for the effective potential in lambda
  phi**4 quantum field theory},
  Phys.\ Rev.\ D {\bf 35}, 1835 (1987).
  
  A.~Duncan and M.~Moshe,
  \textit{Nonperturbative physics from interpolating actions},
  Phys.\ Lett.\ B {\bf 215} (1988) 352.
  
\bibitem{Kneur}
  J.~L.~Kneur, M.~B.~Pinto and R.~O.~Ramos,
  \textit{Asymptotically improved convergence of optimized perturbation theory in the Bose-Einstein condensation problem},
  Phys.\ Rev.\ A {\bf 68}, 043615 (2003) [cond-mat/0207295].

\bibitem{Ricardo}
  R.~L.~S.~Farias, G.~Krein and R.~O.~Ramos,
  \textit{Applicability of the linear delta expansion for the lambda phi**4 field theory at finite temperature in the symmetric and broken phases},
  Phys.\ Rev.\ D {\bf78}, 065046 (2008), arXiv:0809.1449 [hep-ph].

\bibitem{nos}
  M.~C.~B.~Abdalla, J.~A.~Helay\"{e}l-Neto, Daniel L.~Nedel and Carlos R.~Senise Jr.,
  \textit{An extension of the linear delta expansion to superspace},
  Phys.\ Rev.\ D {\bf77}, 125020 (2008), arXiv:0711.0382 [hep-th].
  
\bibitem{PMS}
  P.~M.~Stevenson,
  \textit{Optimized perturbation theory},
  Phys.\ Rev.\ D {\bf 23}, 2916 (1981).
  
\bibitem{Girardello-Grisaru}      
  L.~Girardello and M.~T.~Grisaru,
  \textit{Soft breaking of supersymmetry},
  Nucl.\ Phys.\ B {\bf 194} (1982) 65.

\bibitem{Nibbelink}
  S.~G.~Nibbelink and T.~S.~Nyawelo,
  \textit{Effective action of softly broken supersymmetric theories},
  Phys.\ Rev.\ D {\bf75}, 045002 (2007) [hep-th/0612092].
  
\bibitem{Helayel}
  F.~Feruglio, J.~A.~Helay\"{e}l-Neto and F.~Legovini,
  \textit{Supergraphs extended to broken supersymmetries},
  Nucl.\ Phys. B {\bf249} (1985) 533.
  
\bibitem{nos2}
   M.~C.~B.~Abdalla, J.~A.~Helay\"{e}l-Neto, Daniel L.~Nedel and Carlos R.~Senise Jr.,
   \textit{Work in progress}.
   
\bibitem{Fayet-Iliopoulos}
  P.~Fayet and J.~Iliopoulos,
  \textit{Spontaneously broken supergauge symmetries and Goldstone spinors},
  Phys.\ Lett.\ {\bf51} B (1974) 461.
  
\bibitem{Helayel 2}
  F.~A.~B.~Rabelo de Carvalho, A.~William Smith and J.~A.~Helay\"{e}l-Neto,
  \textit{Superpropagators for broken supersymmetric abelian gauge theories},
  Nucl.\ Phys. B {\bf278} (1986) 309.
  
\end{thebibliography}
\end{document}